\documentclass[12pt]{article}
\usepackage{graphicx,xspace}
\newcommand{\mc     }{\ensuremath{m_{\mathrm{c}}}\xspace}
\newcommand{\lamsq  }{\ensuremath{\Lambda^2}\xspace}
\newcommand{\mzq    }{\ensuremath{{M_{\rm Z}}}\xspace}
\newcommand{\almzq  }{\ensuremath{\al(\mzq)}\xspace}
\newcommand{\epem   }{\ensuremath{{\mbox{e}^+\mbox{e}^-}}\xspace}
\newcommand{\ft     }{\ensuremath{F_{2}^{\gamma}}\xspace}
\newcommand{\fte    }{\ensuremath{F_{2}^{\rm e}}\xspace}
\newcommand{\ftxq   }{\ensuremath{\ft(x,\qsq)}\xspace}
\newcommand{\fl     }{\ensuremath{F_\mathrm{L}^{\gamma}}\xspace}
\newcommand{\flxq   }{\ensuremath{\fl(x,\qsq)}\xspace}
\newcommand{\ftc    }{\ensuremath{F_{2,\mathrm{c}}^{\gamma}}\xspace}
\newcommand{\qsq    }{\ensuremath{Q^{2}}\xspace}
\newcommand{\qnsq   }{\ensuremath{Q_0^{2}}\xspace}
\newcommand{\qzm    }{\ensuremath{\langle \qsq \rangle}\xspace}
\newcommand{\al     }{\ensuremath{\alpha_{s}}\xspace}

\newcommand{\psq    }{\ensuremath{P^{2}}\xspace}
\newcommand{\gev    }{\ensuremath{\mathrm{GeV}}\xspace}
\newcommand{\gevsq  }{\ensuremath{\mathrm{GeV^2}}\xspace}
\newcommand{\ds     }{\ensuremath{D^{*}}\xspace}
\newcommand{\aem    }{\ensuremath{\alpha}\xspace}
\newcommand{\aemsq  }{\ensuremath{\aem^2}\xspace}
\newcommand{\xe     }{\ensuremath{x_{\rm e}}\xspace}
\newcommand{\flux   }{\ensuremath{f_{\gamma}(z,\psq)}\xspace}
\begin{document}
{\bf \hfill \large MPI-PhE/2002-14}
\begin{center}
{\bf \boldmath \LARGE Photon and electron structure \\
 from \epem interactions\footnote{Invited talk presented at the ICHEP02 
 conference, Amsterdam, 25 July 2002.}}
\end{center}
 Richard Nisius, Max-Planck-Institut f\"ur Physik 
 (Werner-Heisenberg-Institut),\\
 F\"ohringer Ring 6, D-80805 M\"unchen, Germany, 
 {\tt E-mail:nisius@mppmu.mpg.de}.
\begin{abstract}
 The status of the measurements and the theoretical developments 
 concerning the hadronic structure of the photon are briefly summarised.
\end{abstract}
%
%
\section{Introduction}
\label{intro}
 For more than 20 years measurements of photon structure functions give deep
 insight into the rich structure of a fundamental gauge boson, the photon.
 A recent review on this subject can be found in~\cite{NIS-9904}.
 \par
\begin{figure}[htb]
\begin{center}
{\includegraphics[width=0.5\linewidth]{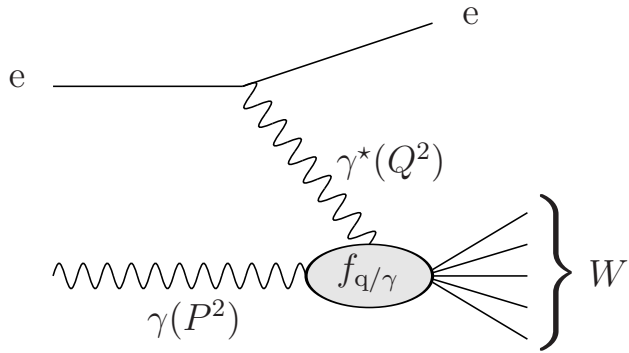}}
\caption{{\it Deep-inelastic electron photon scattering.}
        }\label{fig01}
\end{center}
\end{figure}

 The main idea is that by measuring the differential cross-section
%
 \begin{eqnarray*}
  \frac{d^2\sigma_{{\rm e}\gamma\rightarrow {\rm e} X}}{dxdQ^2}
 &=& \frac{2\pi\aemsq}{x\,Q^{4}}
     \left[\left( 1+(1-y)^2\right) \ftxq - y^{2} \flxq\right]
 \end{eqnarray*}
%
 one obtains the photon structure function \ft, see Figure~\ref{fig01} for
 an illustration.
 Here \qsq and \psq are the absolute values of the four momentum squared 
 of the virtual and quasi-real photons, with $\psq \ll\qsq$. 
 The symbols $x$ and $y$ denote the usual dimensionless variables of 
 deep-inelastic scattering, and \aem is the fine structure constant.
 The flux of the incoming photons, \flux, where $z$ is the fraction of the 
 electron energy carried by the photon, is usually taken from the equivalent
 photon approximation, EPA.
 In leading order, the structure function \ft is proportional to the parton 
 content of the photon and therefore reveals the structure of the photon.
 \par
%
%
 In the region of small $y$ studied ($y\ll 1$), the contribution
 of the term proportional to \flxq is small, and is usually neglected.
 Because the energy of the quasi-real photon is not known, $x$ has to be 
 derived by measuring the invariant mass of the hadronic final state $X$, 
 which is a source of significant uncertainties, and makes measurements of 
 \ft mainly limited by the systematic error, except for large values of \qsq.
 \par
 At this conference new measurements of the hadronic structure function 
 \ft and its charm component \ftc have been presented.
 They are discussed, together with the most recent fits to the \ft data
 and the prospects for measurements of \ft at a future Linear Collider.
 In addition, an attempt by DELPHI is presented to investigate the 
 photon structure by measuring the electron structure function \fte.
%
%
\section{Photon structure function}
\label{f2had}
\begin{figure}[htb]
\begin{center}
{\includegraphics[width=0.7\linewidth]{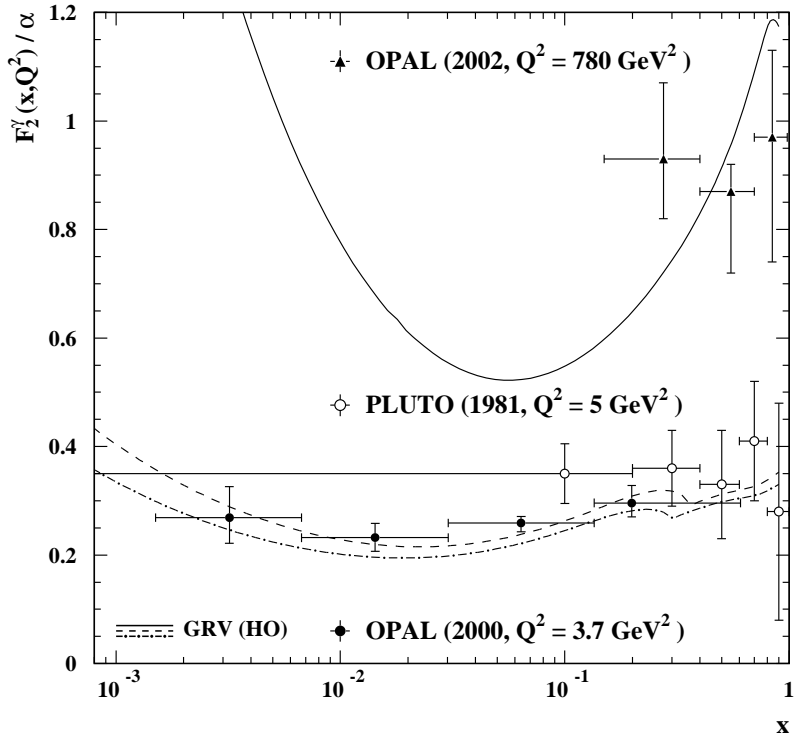}}
\caption{{\it The improvements in \ft at LEP.}
        }\label{fig02}
\end{center}
\end{figure}

 The improvement in the measurement of \ft since the first result
 by PLUTO in 1981 is quite impressive, see Figure~\ref{fig02}.
 The analysis of the LEP data has extended the kinematic coverage by about 
 two orders in magnitude, both to larger \qsq and to smaller $x$.
 In addition, due to continuous improvements of the analyses and a LEP 
 combined effort to obtain a better description of the data by the Monte
 Carlo models, the precision of the measurements has been improved
 considerably.
 \par
%
\begin{figure}[htb]
\begin{center}
{\includegraphics[width=0.7\linewidth]{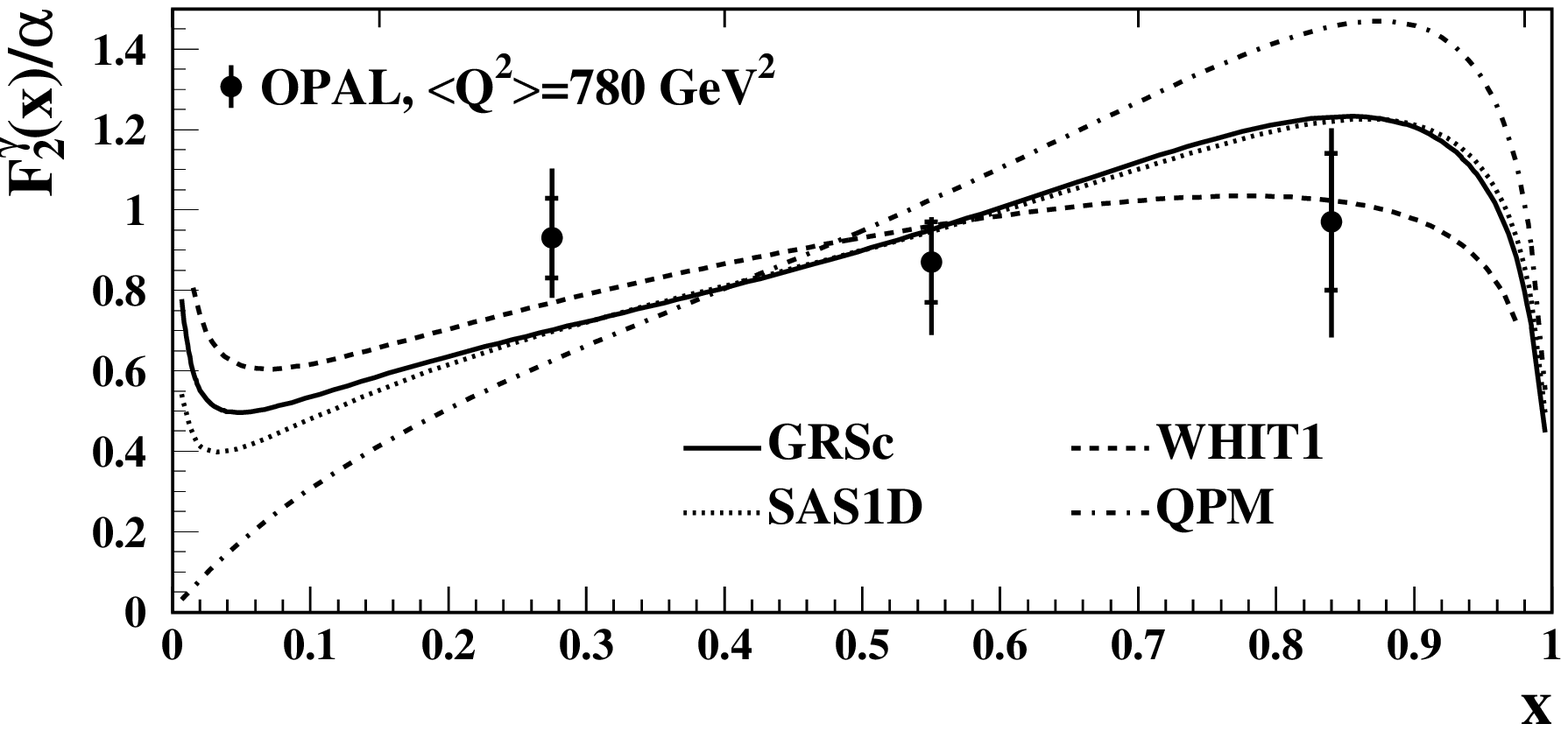}}
\caption{{\it The measurement of \ft at high \qsq.}
        }\label{fig03}
\end{center}
\end{figure}

%
 For this conference the final OPAL result for the measurement of the 
 hadronic structure function \ft at high \qsq has been available.
 This measurement is based on the complete LEP2 data and extends
 the measurement of \ft to $\qzm = 780$~\gevsq, the largest scale 
 ever probed.
 As can be seen from Figure~\ref{fig03} the measured \ft is rather flat,
 and, within errors, the parameterisations from GRSc~\cite{GLU-9902},
 SaS1D~\cite{SCH-9501} and WHIT~\cite{HAG-9501} are in agreement 
 with the data.
 \par
%
%
 The already available preliminary result from DELPHI~\cite{TIA-0101}
 at a slightly lower \qzm basically shows the same trend.
 Since the measurement at high \qsq is mainly limited by the statistical error,
 it is very desirable to combine the OPAL result with the measurements to come 
 from the other LEP experiments.
 To facilitate the combination, the analyses should be performed for the 
 same bins in $x$ and \qsq.
 \par
 Also the investigation of the evolution of \ft with \qsq in ranges of $x$ 
 has been continued using the LEP2 data.
%
\begin{figure}[htb]
\begin{center}
{\includegraphics[width=0.7\linewidth]{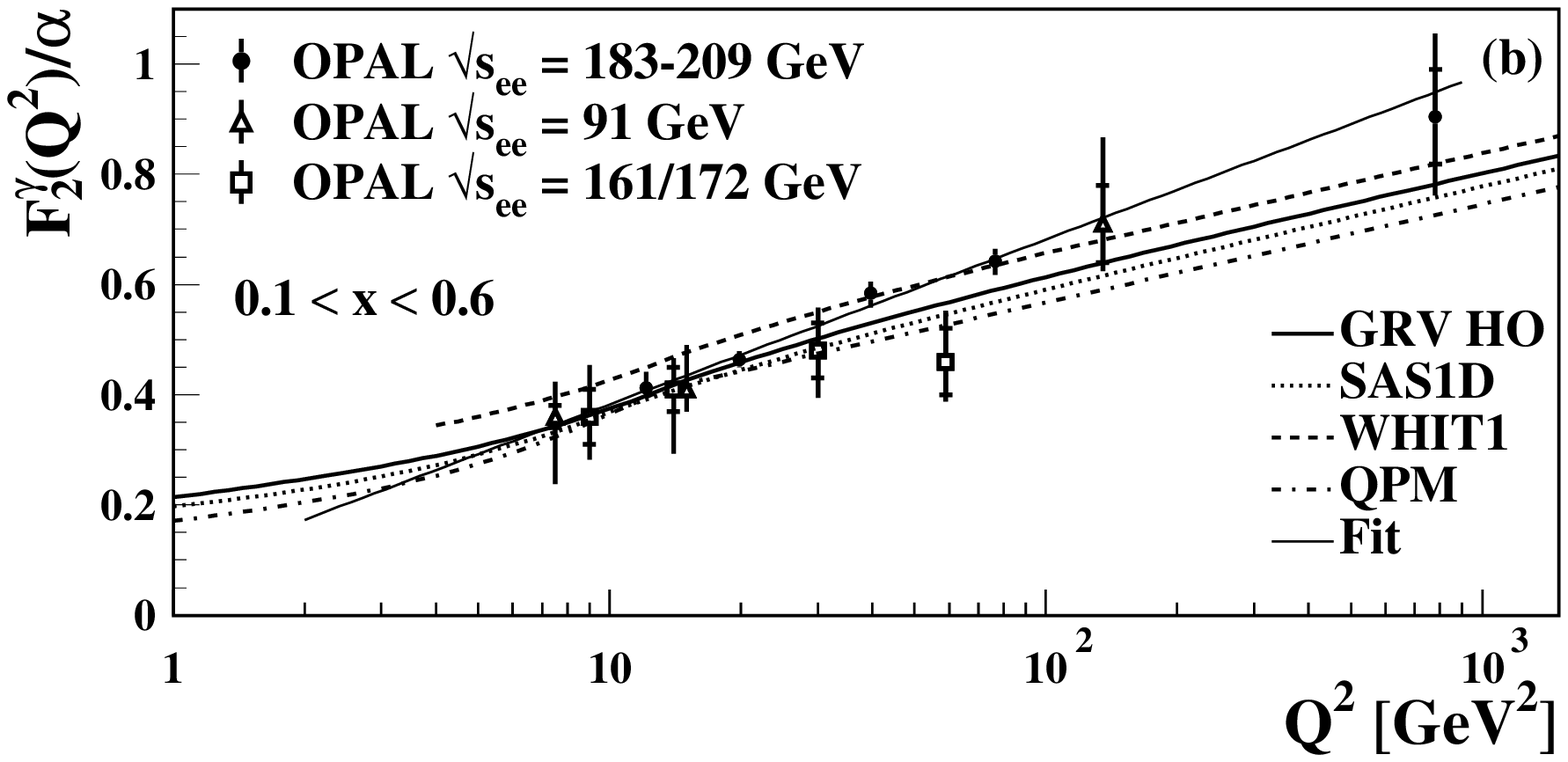}}
\caption{{\it The \qsq evolution of \ft from OPAL.}
        }\label{fig04}
\end{center}
\end{figure}
%
 For medium values of $x$, the precision of the OPAL results based on 
 LEP1 data has been improved considerably by using the large luminosity
 available at LEP2 energies.
 With the present level of precision the data start to
 challenge the existing parameterisations of \ft.
 Given this, several theoretical as well as experimental issues have to
 be addressed in more detail when interpreting the data.
 Examples are the suppression of \ft with \psq and radiative corrections 
 to the deep-inelastic scattering process.
 A summary of the present status of all measurements of the \qsq evolution 
 of \ft is shown in Figure~\ref{fig05}.
 \par
%
\begin{figure}[htb]
\begin{center}
{\includegraphics[width=0.7\linewidth]{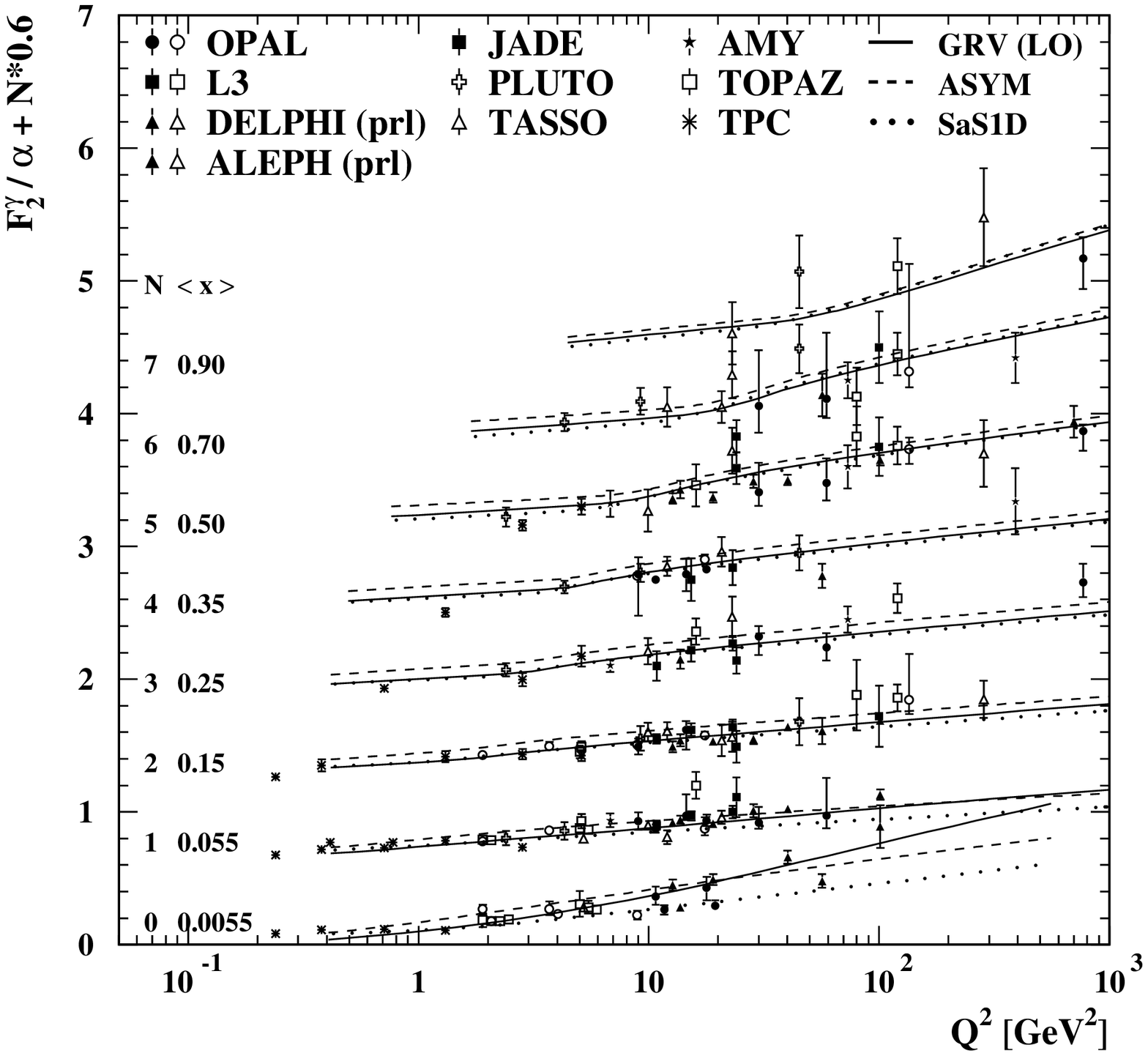}}
\caption{{\it The world data on \ft as a function of \qsq in bins of $x$.}
        }\label{fig05}
\end{center}
\end{figure}
%
%
%
\section{Fits to \ft data}
\label{fit}
 There have been recent fits~\cite{ALB-0201} to \ft based on all
 available data, except for the TPC/2$\gamma$ results.
 To facilitate the analysis, some simplifications are made in the treatment
 of the experimental results. 
 In the analysis the correlation matrix of the various points is used if it 
 is provided by the experiment. However, the systematic errors are treated as
 uncorrelated, the \psq effect and the radiative corrections are neglected,
 and finally, in case of asymmetric errors, the data points are moved to the 
 central value and symmetric error are assumed.
 Given the precision of the data mentioned above, this procedure should 
 most likely be improved for future analyses.
 The fits are done in a fixed flavour scheme with $uds$ as active flavours
 and charm treated as a Bethe-Heitler contribution with $\mc=1.5\pm 0.1$~\gev.
 \par
 Two different types of fits are performed in leading and next-to-leading 
 order, NLO, using the DIS$_\gamma$ and $\scriptstyle\overline{\rm MS}$ 
 schemes.
 In the first fit the hadron-like part of the photon structure is neglected,
 and only the point-like part, which is evolved from a starting scale of 
 $\qnsq=\lamsq$, is taken into account.
 Consequently the only free parameter of the fit is \almzq.
 Since the hadron-like part dominates at low $x$ and \qsq, only the 
 data in the region $x>0.45$ and $\qsq>59~\gevsq$ are used in this fit.
 The second fit uses all data, takes into account both components and 
 fits for $(N,\,\alpha,\,\beta,\,\al,\,\qnsq)$
 using the assumptions, $uds(\qnsq)=N x^\alpha(1-x)^\beta$ and $g(\qnsq)=0$.
 Both types of fits give a reasonable description of the data.
 Within the assumptions made, the quoted theoretical precision on \almzq is 
 about 3$\%$ and predicted to shrink to less than 2$\%$ for \qsq values
 larger than 300~\gevsq.
%
%
\section{Prospects for \ft measurements}
\label{prosper}
\begin{figure}[htb]
\begin{center}
{\includegraphics[width=0.67\linewidth]{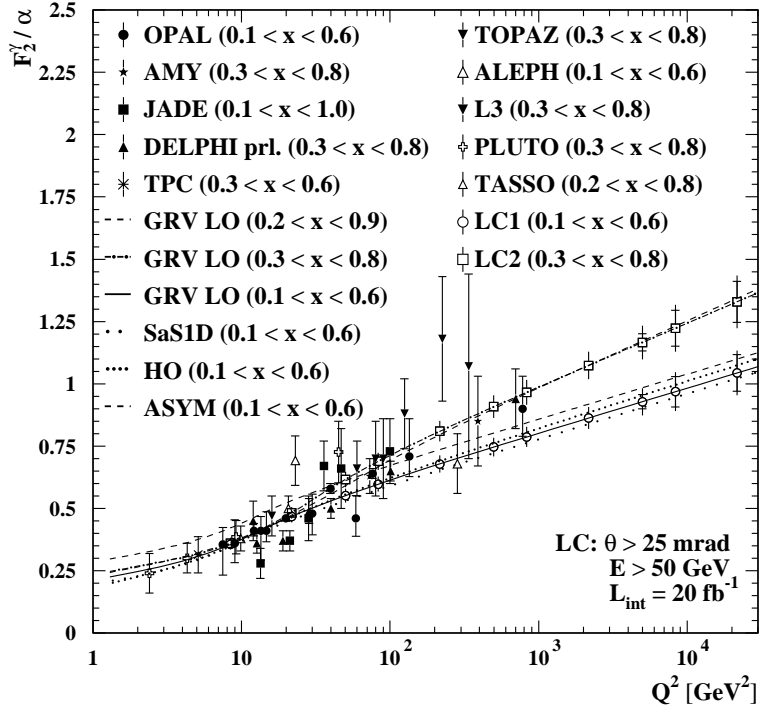}}
\caption{{\it The expected measurement of \ft at a future Linear Collider.}
        }\label{fig06}
\end{center}
\end{figure}

 The prospects of future investigations of the photon structure 
 in the context of the planned Linear Collider programme are very 
 promising.
 The \epem Linear Collider will extend the available phase space, as shown
 in Figure~\ref{fig06}, for the measurement of the \qsq evolution of \ft
 at medium $x$, see~\cite{NIS-9904}.
%
%
 The higher beam energy and luminosity compared to LEP also allows
 for the investigations of novel features like the measurement of
 the flavour decomposition of \ft by exploring the exchange of
 $W$ and $Z$ bosons~\cite{GEH-9901}.
%
%
\section{The charm contribution to \boldmath\ft}
\label{f2ch}
\begin{figure}[htb]
\begin{center}
{\includegraphics[width=0.7\linewidth]{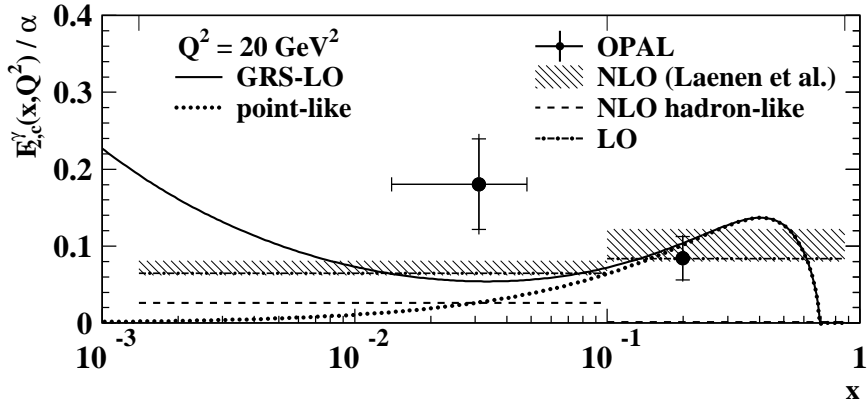}}
\caption{{\it The measurement of \ftc from OPAL.}
        }\label{fig07}
\end{center}
\end{figure}

 The final OPAL result~\cite{OPALPR354} has been presented of the 
 measurement of the charm component \ftc using \ds mesons to identify 
 charm quarks.
 Compared to the first OPAL result on \ftc, this analysis is based on 
 improved Monte Carlo models and higher statistics, leading to a more
 precise measurement presented in Figure~\ref{fig07}.
%
%
 In a similar way to the structure function for light quarks, \ftc receives 
 contributions from the point-like and the hadron-like components of the 
 photon structure. 
 These two contributions are predicted~\cite{LAENEN} to have 
 different dependences on $x$, with the hadron-like component dominating 
 at very low values of $x$ and the point-like part accounting for most 
 of \ftc at $x>0.1$.
 \par
 For $x>0.1$ the OPAL measurement is described by perturbative QCD 
 at next-to-leading order.
 For $x<0.1$ the measurement is poorly described by the NLO prediction
 using the point-like component alone, and therefore the measurement
 suggests a non-zero hadron-like component of \ftc.
 Increased statistics and a better understanding of the dynamics 
 for $x<0.1$ are needed to get a more precise result in this region.
 Also here, to increase the statistics, it would be advantageous to combine 
 the data from all four LEP experiments taken in the same phase space.
%
%
\section{Electron structure function}
\label{elec}
 For this conference there has been an attempt by DELPHI to access the 
 photon structure by measuring the electron structure function.
 In this analysis, instead of measuring the electron-photon scattering
 cross-section by utilising the EPA to account for the flux of the 
 quasi-real photons, the cross-section of electron-electron scattering is 
 studied as functions of \qsq and \xe, the fractional momentum of the
 parton with respect to the electron.
 This quantity is related to $x$ by $\xe=zx$.
 The main advantage of this approach is experimental, i.e.~the incoming 
 particle probed by the virtual photon is the electron and not the photon.
 This means that its energy is known, and therefore \xe can be obtained 
 without measuring $W$.
 However, there is also a disadvantage in this measurement.
 The photon structure is obscured because, e.g.~the region of low values of
 \xe receives contributions from both, the region of large momentum fraction
 $x$ and low scaled photon energy $z$, and the region of small $x$ and 
 large $z$.
 The measurement of \fte is performed with a precisions 
 of about 3-20$\%$ both for the statistical and the systematic error.
 So far, no radiative corrections and no bin-centre corrections are applied.
 The preliminary DELPHI result is consistent with several existing 
 parameterisations of \fte, obtained from \ft and \flux by convolution.
 The usual exception is the LAC1 parameterisation, which is disfavoured.
 This investigation serves as a valuable cross-check of the \ft measurements,
 but does not give more insight into the photon structure.
%
%
\section{Conclusion}
\label{concl}
 Given the large statistics available at LEP2 energies, the region of 
 phase space covered in the investigations of the structure of the photon 
 is constantly increasing.
 Despite these large luminosities, for some of the measurements the results 
 are still limited by the statistical error and a combination of the results
 from several experiments is desirable.
 This is particularly true for the measurement of \ft at large \qsq and 
 the determination of \ftc.
 For the first time, the high precision data from LEP have been used in
 NLO fits to \ft results. 
 With an improved treatment of the experimental data, and the inclusion of the 
 jet data from HERA to better constrain the gluon distribution in the photon,
 there are good prospects to achieve the first parametrisation of the parton 
 distributions of the photon based on LEP and HERA data in the near future.
%
%

%
%
\end{document}